\documentclass[journal, 10pt, twocolumn]{IEEEtran} %
\usepackage{cite} 
\usepackage{tikz}

\usepackage{amsmath,amsfonts,amssymb,amsthm}

\usepackage{subcaption} 
\usepackage{hyperref}
\usepackage{caption}
\usepackage{gensymb} 
\usepackage{xspace}
\usepackage{mathtools} 
\usepackage{cuted}
\usepackage{stfloats}
\usepackage{multirow}
\usepackage{balance} 
\usepackage[breakable, skins]{tcolorbox}
\tcbset{arc=0mm,size=fbox,attach title to upper={\ ---\ },coltitle=black}

\newcommand{\B}{\mathcal{B}}

\newcommand{\U}{\mathcal{U}}

\newcommand{\snir}{\gamma}
\newcommand{\rate}{C}
\newcommand{\fdp}{\text{fdp}\xspace}
\newcommand{\fsp}{\text{fsp}\xspace}
\newcommand{\FDP}{\text{FDP}\xspace}
\newcommand{\FSP}{\text{FSP}\xspace}
\newcommand{\KPN}{MNO$_{1}$\xspace}
\newcommand{\TMobile}{MNO$_{2}$\xspace}
\newcommand{\Vodafone}{MNO$_{3}$\xspace}
\newcommand{\effectiveRes}{\xi}
\newcommand{\ploss}{\mathcal{L}}

\newcommand{\newInfo}[1]{\textcolor{black}{#1}}

\DeclareMathOperator*{\argmax}{arg\,max} 

\begin{document}
\title{On the resilience of cellular networks: how can national roaming help?}

\author{Lotte Weedage, Syllas R. C. Magalh\~{a}es, Clara Stegehuis and Suzan Bayhan
\thanks{Authors are with the Faculty
of Electrical Engineering, Mathematics and Computer Science~(EEMCS), University of Twente, The Netherlands,
Corresponding author's e-mail: l.weedage@utwente.nl}
}


\maketitle

\begin{abstract}

Cellular networks have become one of the critical infrastructures, as many services depend increasingly on wireless connectivity. 
Therefore, it is important to quantify the resilience of existing cellular network infrastructures against potential risks, ranging from natural disasters to security attacks, that might occur with a low probability but can lead to severe disruption of the services. 
In this paper, we combine models with public data from national bodies on mobile network operator~(MNO) infrastructures, population distribution, and urbanity level to assess the coverage and capacity of a cellular network at a country scale. Our analysis offers insights on the potential weak points that need improvement to ensure a low fraction of disconnected population~(FDP) and high fraction of satisfied population~(FSP).  
As a resilience improvement approach, we investigate in which regions and to what extent each MNO can benefit from infrastructure sharing or \textit{national roaming}, i.e., all MNOs act as a single national operator. 
As our case study, we focus on Dutch cellular infrastructure and model risks as random failures and correlated failures in a geographic region. 
Our analysis shows that there is a wide performance difference across MNOs and geographic regions in terms of FDP and FSP. However, national roaming consistently offers significant benefits in some regions, e.g., up to 13\% improvement in FDP and up to 55\% in FSP when the networks function without any failures. We then show that a similar performance improvement can be obtained by partial implementation of national roaming.
\end{abstract}


\IEEEpeerreviewmaketitle

\section{Introduction}\label{sec:introduction}

Cellular networks play a key role in today's communications, as many services depend on the proper functioning of these infrastructures.    
However, they can be vulnerable to failures resulting from various sources such as 
large-scale natural disasters including earthquakes~\cite{mayer-eq-pam21} and wildfires~\cite{anderson-wildfires-imc2020}, cyberattacks on the network infrastructure~\cite{vodafoneportugal, mukherjee2014network}, or regional power shortages~\cite{reutersnews}. 
{These events will either affect certain regions, such as earthquake areas \cite{mayer-eq-pam21}, or can be randomly spread~(e.g., hardware-related failures).}
Indeed, the functioning of cellular networks becomes even more important during such failure events, e.g., for rescue and recovery in the aftermath of disasters. The key question then is: \textit{what is the coverage and capacity that a mobile network operator~(MNO) can provide, given some links or network nodes do not function?}%
While the resilience literature is broad in other areas of critical infrastructures, to the best of our knowledge, there are only few studies on quantifying a cellular network's resilience at a national scale such as \cite{griffith2015measuring} and \cite{yanPredictiveImpact}, the former defining resilience as ``the maximum number of sites that can fail before the performance metric of interest falls below a minimum acceptable threshold", and the latter using the number of served users as the resilience metric. 
Since both ensuring coverage
and satisfying quality of service~(QoS) are important, we use \textit{fraction of disconnected population}~(FDP) and \textit{fraction of satisfied population}~(FSP) \newInfo{considering data services} to quantify the resilience of an MNO. 

Combining cell tower data with population density statistics as well as urbanity levels in the Netherlands, we investigate the current state of the Dutch MNOs\footnote{Our analysis can be extended to other countries by providing the data presented in Sec.\ref{sec:data} on the MNO infrastructures, cities, urbanity of each region, and population data of the corresponding countries.} and then study their resilience to (i) random failures which could occur due to human errors and (ii) failures confined to a certain geographical region occurring due to disasters. 
The insights from our analysis can help to improve the MNO infrastructures to absorb crises or to recover quickly from their effects.

While there are various ways to improve the resilience of a cellular network, one approach is \textit{national roaming} which facilitates MNOs to use the infrastructure of each other when needed, e.g., in case its own infrastructure is not functional or does not suffice to serve with the required service levels. 
{For increasing \newInfo{data} coverage, national roaming has already been widely adopted in countries such as India~\cite{indiaRoaming13} where MNOs often operate in certain parts of the country or where infrastructure is not as ubiquitous. In 2022, the Federal Communications Commission (FCC) in the US has introduced \textit{roaming under disasters}~(RuD) based on bilateral agreements between MNOs that requires MNOs to serve each other's users in case of service disruptions due to the infrastructure damage during disasters and emergencies~\cite{fcc2022}. Moreover, in 2022 after national infrastructures are damaged considerably due to the Russian invasion, MNOs in Ukraine implemented national roaming~\cite{ukraineNR2022}.} 
Prior studies such as \cite{ENISA2010} advocate national roaming for more resilient cellular infrastructures, and studies such as \cite{panchal2013mobile} investigate different modes of MNO network sharing. \textit{However, to the best of our knowledge, there is no study quantifying the gains in coverage, capacity, and resilience facilitated by national roaming.}  
Toward filling this gap, we quantify in this paper the gains when MNOs work together to serve all the citizens as a single national operator. 
{Furthermore, to provide some insights to the MNOs or to the national telecommunication regulatory bodies on in which areas and which technologies they could prioritize such roaming agreements, we analyze implementation of roaming in a more restricted way such as in only urban areas or rural areas, or considering a certain technology such as 4G or 5G.}

To summarize, our goal in this paper is to address the following questions: 
\begin{itemize}
\item What is current coverage and capacity of MNOs in the Netherlands? Are there regional differences~(e.g, across cities) or differences among the MNOs? 
\item What would be the performance gain if all MNOs act as a single national operator and in which regions will this provide the highest gains? 
\item How resilient are MNOs to various types of failures and how much can national roaming help to ensure resilience of the MNOs against failures?
\item 
{Which technologies (3G, 4G, 5G) or which areas (urban vs. rural) should MNOs consider first implementing such roaming agreements for ensuring the highest benefits in terms of coverage and user satisfaction}  \newInfo{for data services?}

\end{itemize}
In a nutshell, our contributions are as follows:
\begin{itemize}
   \item We provide an approach to assess the resilience of a cellular network using  public data and models on coverage and capacity. To reflect the effect of disruptions on the citizens, we use FDP and FSP as our metrics.  
   \item Using publicly-available data from national bodies in the Netherlands, we assess the current state of the Dutch cellular networks on both province and municipality level.
   \item We show that national roaming leads to significant benefits {in some regions} for both FSP and FDP while there is a large performance difference across MNOs and geographic regions. {These areas with high performance gains from national roaming could be considered first by  MNOs to enter such roaming agreements. Moreover, we provide an analysis of roaming implemented only for certain cellular technologies, e.g., 3G and 5G, and in urban vs. rural areas 
   in case MNOs prefer national roaming in a more limited way rather than sharing their all network infrastructures.}
   \item We model two risk scenarios to investigate the resilience of the Dutch MNOs with and without national roaming. Our analysis suggests that MNOs are resilient against isolated failures owing to high \newInfo{base station~(BS)} density. That is, FDP  remains roughly the same. On the contrary, FSP  decreases due to the increased number of users served by the surviving BSs. Meanwhile, the impact of correlated failures is more drastic due to BSs  in the same region becoming dysfunctional simultaneously. {To ensure resilience in such cases, alternative approaches are paramount, e.g., aerial connectivity or cells on wheels.}
   \end{itemize}

The rest of the paper is organized as follows. First, Section~\ref{sec:related} provides an overview of the related work on resilience metrics and analysis of communication networks and national roaming. Then, Section~\ref{sec:system_model} introduces the considered system model which is followed by Section~\ref{sec:res_metrics} that presents the definition of the metrics used in our analysis. Next, Section~\ref{sec:perfeval} presents a case study of Dutch cellular networks\footnote{\newInfo{We share the code and data publicly under  \url{https://github.com/lweedage/disaster-resilience}. }} and publicly-available data, i.e., on MNOs, population statistics, and urbanity levels of different areas. 
Section~\ref{sec:perf-failures} also considers the failures and how they will affect the MNOs. Finally, Section~\ref{sec:discussion} discusses the limitations of our work and Section~\ref{sec:conclusion} draws  conclusions.

\section{Related Work}\label{sec:related}
This section discusses the related work on  \textit{resilience metrics and analysis} and   \textit{national roaming/infrastructure sharing}.

\smallskip
\noindent \textbf{Resilience metrics and analysis:} The resilience of networks have been investigated in different contexts, such as in traffic networks~\cite{ganin2017}, power grids~\cite{kim2017network}, ecological networks~\cite{kharrazi2016evaluating} and supply chain networks~\cite{zhao2011analyzing}. 
The resilience of the network has several aspects: whether the network is operational, and whether the service it offers is acceptable. To represent this distinction, metrics are categorized into two as \textit{topological} and \textit{functional metrics}~\cite{rak2020guide}. Topological metrics, 
represent the status of the underlying network's connectivity and network paths, e.g., 
relative size of the largest connected component, average two terminal reliability, average path length, motifs~\cite{dey2019} and spectral metrics~\cite{alenazi2015evaluation}. 
While these topological metrics provide useful insights about the existing/surviving infrastructure, they fall short of assessing the satisfaction of the served users and applications. As a remedy, functional metrics including objective~(e.g., link stress and node load~\cite{rak2020guide}) and subjective metrics~(e.g., mean-opinion-score or other quality-of-experience metrics~\cite{natalino2020functional}) aim at assessing to what extent the network can satisfy its users' expectations. 

Note that functional metrics need to reflect the application's requirements, e.g., a reliability requiring application will be assessed by packet loss ratio. Some studies focus on static metrics, e.g., performance after the disruptive event, while others measure the system performance over a period of time~\cite{sun2020resilience, lin2020}. Similar to \cite{liang2023holistic}, our study focuses on static resilience using functional metrics, where we investigate the loss in quality of service immediately after a disruptive event.   

When it comes to analyzing the resilience of a network, the closest studies to ours are \cite{griffith2015measuring, yanPredictiveImpact}. Yan et al.\cite{yanPredictiveImpact} propose the metric Tower Outage Impact Predictor~(TOIP) to quantify the impact of a failure of a cell tower on the number of served cellular users. Due to the dense deployment of cell towers, a user might not perceive the failure of a cell-tower as it will be served immediately by another BS in the proximity.  
Using the data from an operational MNO, the authors propose a prediction scheme to estimate the number of users who would be affected by service outages of one or more BSs in the network. Our work differs from \cite{yanPredictiveImpact} in that we combine cellular network coverage and capacity models with the real-world data on the cellular network infrastructure to quantify the impact of failures on the user's connectivity and satisfaction performance. Moreover, our focus is also on the potential resilience improvement due to infrastructure sharing among MNOs in the same country. 
\textcolor{black}{Contrary to~\cite{griffith2015measuring}, our work investigates both coverage as service quality as metrics, and combines models with several data sources to assess resilience.}

Authors of \cite{smallcells} consider two kinds of failures that might affect small cells in the Netherlands: power outages and cyberattacks. They investigate a case of power outage in 2015 in Noord-Holland, where after two hours $80\%$ of the BS sites were down. In this case, providers immediately shut down the LTE network to have sufficient power supply for the essential network. Study in \cite{smallcells} proposes to also first deactivate small cells for keeping the emergency services functioning longer.

The resilience of 5G networks is of particular importance, since higher frequencies~(e.g., mmWave  bands) are more prone to errors~\cite{Esposito2018} and links operating at these high bands might be affected by rain drastically, as the rain drops are non-negligible in their size in relation to the wavelength of the mmWave signals~\cite{cinkler2020resilience,hemadeh2017millimeter}. An eight-year measurement study \cite{padmanabhan2019residential} shows that rain might result in internet outages in some regions, and especially wireless links are more prone to these outages.  
\textcolor{black}{Our study differs from these studies in that we combine models and publicly-available data to offer insights on the coverage and capacity performance of MNOs under two failure categories and the potential of national roaming toward mitigating the impact of these failures.}

\textbf{National roaming/infrastructure sharing: }
As resilience can be improved by increasing redundancy and over-provisioning \cite{ENISA2010}, some studies investigated how infrastructure sharing can help from an economic or operations perspective \cite{frisanco2008infrastructure, zheng2017economic}, increase rate coverage probability \cite{jurdi2018modeling, sanguanpuak2018infrastructure}, decrease energy consumption \cite{gambin2019sharing}, or distribute the load evenly over the network \cite{difranceso14}. 
{Next to improved resilience, infrastructure sharing plays a key role in the future of telecommunications to keep the services affordable in developing countries \cite{oughton2022policy, oughton2023policy} and to increase coverage \cite{arakpogun2020threading}.
Therefore, nationwide infrastructure sharing might be difficult to achieve due to possible decrease in operator revenue~\cite{baake2021mobile}.
To mitigate this, one might consider to only share the infrastructure among certain operators, certain radio types~\cite{9121242} or on certain areas (Rural Access Network, \cite{oughton2022policy}). Please refer to \cite{ituTinfrasharing12} for an overview of different modes of infrastructure sharing, e.g., at the core network or radio access network.} {Note that virtualization techniques already implemented by Mobile Virtual Network Operators~\cite{liang2014wireless} can be considered for implementation of the specific national roaming scenario studied in this paper, i.e., a single MVNO having access to all resources of all national MNOs.}

Different from the studies that focus on infrastructure sharing \cite{panchal2013mobile, meddour2011role}, we focus on the special case of \textit{national roaming} where every user can use every existing BS of any MNO, { similar to the approaches in \cite{ENISA2010} and \cite{oughton2023policy}}. 
We investigate the entire network and focus on improving resilience instead of purely on maintaining higher capacity. 
Moreover, since our goal is to quantify the full potential of national roaming, rather than considering national roaming only as a backup implemented in exceptional cases such as post-disasters or for a single MNO, we explore it as a default mode of operation where all MNOs can use each other's network for serving the customers in the most efficient way.
Therefore, this can be considered as the best case scenario where all MNOs are in business agreement 
 { and operating with a shared core network}. 
Note that the economic implications, e.g., settling the roaming costs, are beyond the scope of our paper. {Please refer to \cite{hou2020economics} for a thorough discussion on the economic aspects of infrastructure sharing.}

\section{System Model}\label{sec:system_model}

 \begin{figure}
   \centering
    \includegraphics[width = 0.45\textwidth]{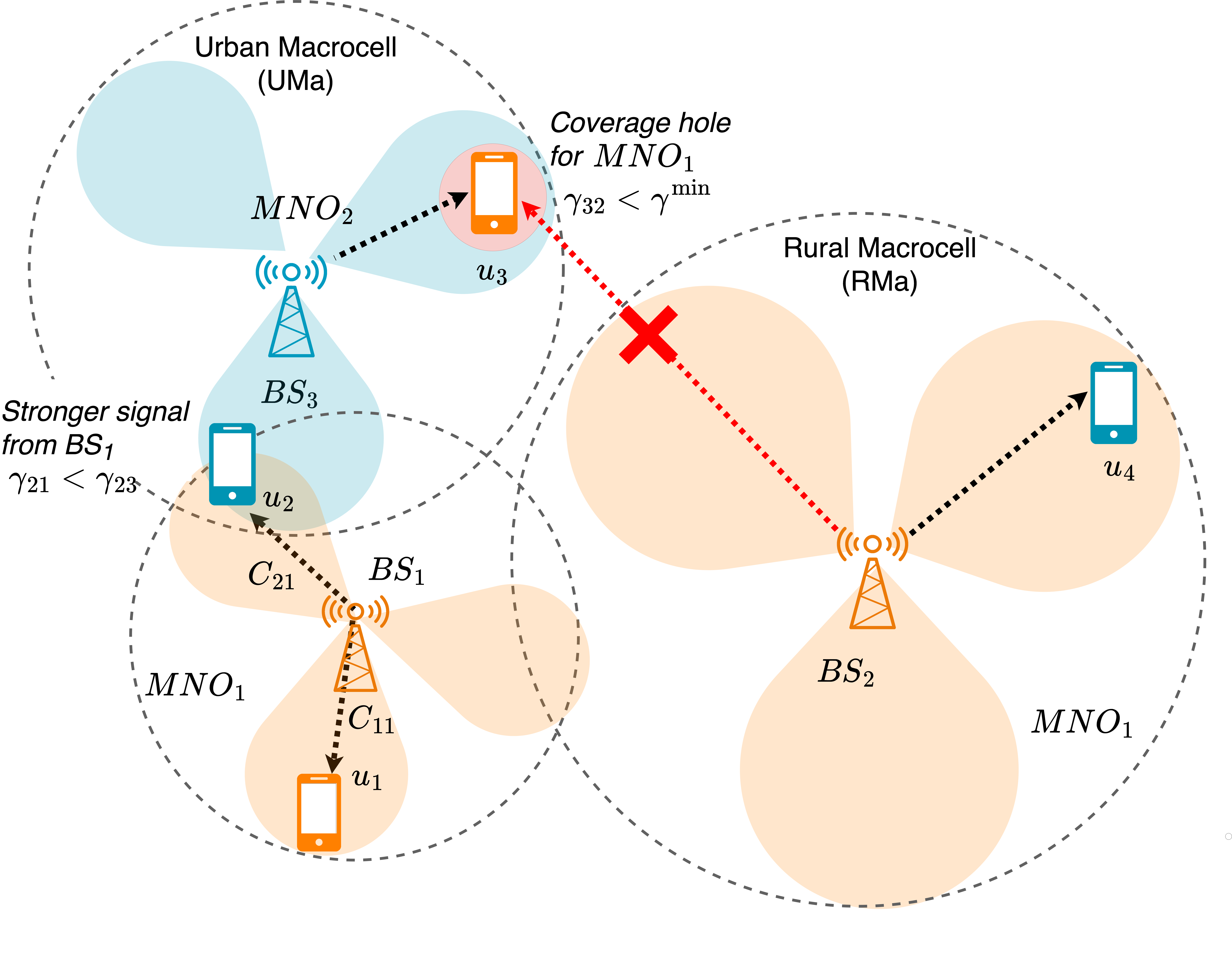}
    \caption{An illustration of the considered system model with two MNOs, MNO$_{\textrm{1}}$ (with subscribers $u_1$ and $u_3$) and MNO$_{\textrm{2}}$ (with subscribers $u_2$ and $u_4$). MNO$_1$'s coverage hole can be filled with MNO$_2$'s base station BS$_3$ to serve $u_3$. Moreover, to have a stronger signal, $u_2$ can connect to MNO$_1$'s BS$_1$. } 
    \label{fig:sysmodel}
\end{figure}
We consider a cellular network as in Fig.\ref{fig:sysmodel} consisting of a set of BSs denoted by $\B$ which operate at sub-6\,GHz bands. The network serves a set of users denoted by $\U$ and each user $u_i$ has a minimum rate requirement denoted by $C_i^{\textrm{min}}$ bits per second for its application to sustain a satisfying user experience.
If there is a link between user $u_i \in \U$ and BS$_j \in \B$, we denote this channel by $\ell_{ij}$. 

We consider two types of cell sites following the models in 3GPP~\cite{3gpp38901}: rural and urban macrocells. We model path loss according to the 3GPP TR 38.901 specification~\cite{3gpp38901} which defines path loss models for each listed cell type. For each case, line-of-sight~(LoS) and non-LoS path loss probabilities and models are defined in \cite[Table 7.4.1-1]{3gpp38901}. We denote the path loss at a receiver $u_i$ located at $r_{ij}$ meters away from the transmitter BS$_j$ by $\ploss(r_{ij})$. 

We consider that a BS has a three-sector antenna and we adopt the 3GPP antenna gain model for these three-sector antennas~\cite{3gpp38901}. The horizontal antenna gain $A_H$ is defined as:
\begin{align}
    A_H(\phi)[\text{dB}] &= - \min\left\{12 \left(\frac{\phi}{\phi_{\text{3dB}}}\right)^2, 20\right\},\label{eq:AH}
\end{align}
for horizontal misalignment angle $\phi$ (in degrees). 
The angle $\phi_{\text{3dB}}$ denotes the horizontal 3dB beamwidth. For three sector antennas, we assume $\phi_{\text{3dB}}=65\degree$\cite{rebato2019stochastic}. Thus, for user $u_i$ that connects to BS$_j$, the antenna gain is:
\begin{align}
    G_{ij}[\text{dB}] = G_{\text{max}} +A_H(\phi), \label{eq:antenna_gain_3GPP}
\end{align}
where $G_{\text{max}} = 20$ dB, which is the maximum attenuation. We assume users have omnidirectional antennas, and hence the receiver antenna gain is 0\,dB.

Now, let us denote the transmission power of BS$_j$ by $P_j$ and total noise over the transmission band by $N_{\text{tot}}$ which represents the sum of thermal noise power and noise figure of the receiver~\cite{halperin2013simplifying}. 
Then, the signal-to-interference-and-noise-ratio~(SINR) $\gamma_{ij}$ for $\ell_{ij}$ is defined as:
\begin{align}
    \gamma_{ij} = \frac{P_{j} G_{ij} \ploss(r_{ij})^{-1}}{N_{\text{tot}} +  I_{ij}}, \label{eq:snr}
\end{align}
where $G_{ij}$ denotes the antenna gain and $I_{ij}$ denotes the interference perceived by $u_i$ when it receives its downlink traffic from BS$_j$. %
Formally, $I_{ij}$ is defined as:
\begin{align}
    I_{ij} &= \sum_{m \in \B^j}P_{m} G_{im} \ploss(r_{im})^{-1}, \label{eq:interference}
\end{align}
where $\B^j$ is the set of BSs that operate on the same frequency as BS$_j$ and are within a radius of certain distance $r_\text{max}$ from BS$_j$.
We assume that the closest three BSs to BS$_j$ implement interference coordination schemes~\cite{hamzaicicsurvey13}. Hence, they do not interfere with BS$_j$.
For a signal to be decodable with the lowest modulation scheme, we assume that the required minimum SINR is $\snir^{\textrm{min}}$.

Finally, we can derive the channel capacity of user $u_i$ if it is connected to BS$_j$ (denoted by $C_{ij}$, Fig.~\ref{fig:sysmodel}).
We assume that the BS applies time-sharing among its served users. 
Then, the user's \textit{effective bandwidth}\footnote{We use the term \textit{effective bandwidth} to reflect the bandwidth the user will experience if it is served a certain amount of time in the available bandwidth, i.e., $\effectiveRes_{ij}$. 
Hence, the effective bandwidth is $\effectiveRes_{ij} W_j$.} will be only a fraction $\effectiveRes_{ij}$ of the total bandwidth $W_j$. 
Consequently, we can calculate the maximum achievable throughput $C_{ij}$ as:
\begin{align}
   C_{ij} = \effectiveRes_{ij}W_j\log_2(1 + \gamma_{ij}).\label{eq:channel_capacity}
\end{align}

Now, let us discuss how to determine  $\effectiveRes_{ij}$. 
Ideally, each user should be served for a minimum amount of time such that its rate requirement $C^\text{min}_i$ is met. Let us denote this minimum  effective bandwidth by $W^{\textrm{min}}_{ij}$ that is needed to ensure 
$C^\text{min}_i$. We can calculate $W^{\textrm{min}}_{ij}$  as follows:
\begin{align}
W^{\textrm{min}}_{ij} = \frac{C^\text{min}_i}{\log_2(1 + \gamma_{ij})}.
\end{align}
Since the total requested effective bandwidth by all users might exceed the available bandwidth $W_j$, for the sake of fairness among all users, each user is assigned effective bandwidth proportionally to its need $W^{\textrm{min}}_{ij}$. Hence, we set $\effectiveRes_{ij}$ as follows:
\begin{align}
    \effectiveRes_{ij} = \frac{W^{\textrm{min}}_{ij}}{\sum_{k\in\U^j} W^{\textrm{min}}_{kj}},\label{eq:effective_bandwidth}
\end{align}

where $\mathcal{U}^j$ denotes the set of users connected to BS$_j$. When $\sum_{i\in\U^j} W^{\textrm{min}}_{ij} \leqslant W_j$, the bandwidth is sufficient for all connected users and users  maintain at least their requested rate. %
Otherwise, users are assigned effective bandwidth proportional to their $W^{\textrm{min}}_{ij}$ with respect to the needed bandwidth by all users, i.e.,  $\sum_{i\in\U^j} W^{\textrm{min}}_{ij}$. \newInfo{Note that we consider data services in this paper, hence our metric to represent user satisfaction is based on the user's data rate. However, for other services requiring guaranteed latency or reliability, one would need different metrics for measuring performance and consequently resilience.}

We assume that active users are distributed by a Poisson Point Process with given density according to the population density. As the user association scheme, we assume that a user connects to the BS that offers a high SINR that is above a threshold SINR~($\gamma^{\text{min}}$) with the lowest number of users connected to it. We iteratively connect a randomly-chosen user $u_i\in \U$ to $\text{BS}_{\text{opt}}$ where  $\text{BS}_{\text{opt}}$ is: 
\begin{align}
    \text{BS}_{\text{opt}} = \argmax_{j\in \B}\frac{
    \gamma_{ij}}{D_{\text{BS}_j}},  
    \label{eq:user_association}
\end{align}
where $D_{BS_j}$ is the number of users connected to BS$_j$. 
Finally, we denote by $X=[x_{ij}]$ the association state of $u_i$ with BS$_j$ where $x_{ij}$ yields value 1 if $u_i$ is associated with BS$_j$ and zero otherwise. 

\section{Two metrics to assess network resilience}\label{sec:res_metrics}

As described in Sec.~\ref{sec:related}, resilience can be measured both using topological metrics and functional metrics. However, none of the discussed metrics in Sec.~\ref{sec:related} reflect what resilience implies for the citizens, i.e., customers of an MNO. %
For a cellular network, two properties are key to assess performance: coverage and capacity. For the coverage, we quantify the \textit{fraction of disconnected population} (per province/city) as a performance metric. We will refer to this metric as FDP. We assume that a user is disconnected from the network if its received signal strength is too low to decode the signal with the most robust modulation~(e.g., BPSK). 
Note that this metric is an adapted version of SINR coverage that is typically used for capacity analysis in cellular networks, e.g., \cite{rebato2019stochastic}.

To measure capacity, we will use the throughput that a user maintains. However, since functional metrics describe the level of service that users experience, it depends on their application requirements. 
Indeed, cellular networks can support many applications with vastly different rate requirements, which also impacts the satisfaction level of a user with a given throughput. Therefore, we  quantify the \textit{fraction of satisfied population} (per province/city) and refer to it as FSP. 
A user is \textit{satisfied} if the user is connected to the network~(i.e., it is not in the disconnected population) and the provided throughput to this user is above the minimum rate requirement of the used application. 
Note that applications can be diverse, e.g., URLLC or mMTC services in 5G networks require a low data rate but can assert strict latency or reliability performance \cite{8476595}. 
In this study, we consider only the rate requirements in defining our satisfaction metric.

Now, we can define FDP and FSP formally. Let us denote whether $u_i$ is disconnected from the network by $\delta^{\fdp}_{i}$. We define it as follows:  
\begin{align}
\delta^{\text{fdp}}_{i} = \begin{cases}
    1 & \textrm{if}  \sum\limits_{j\in \B}\snir_{ij}x_{ij} \leqslant \snir^{\textrm{min}} \\
    0 & \textrm{otherwise}.
\end{cases}
\end{align}
Then, we can calculate the network's coverage performance in terms of fraction of disconnected population as: 
\begin{align}
\FDP = \frac{\sum_{i\in|\U|} \delta^{\fdp}_i}{|\U|}.
\end{align}

For FSP, we denote the satisfaction status of $u_i$ by $\delta_i^{\fsp}$ and define it as follows:
\begin{align}
\delta^{\fsp}_{i} = \begin{cases}
    1 & \textrm{if} \sum\limits_{j\in |\B|}\rate^s_{ij}x_{ij} \geqslant \rate^{\textrm{min}}_i { \textrm{ and } } {\color{black}\delta^{\text{fdp}}_{i} = 0}\label{eq:fsp_1} \\
    0 & \textrm{otherwise}.
\end{cases}
\end{align}
Similarly, the FSP can be defined as: $\FSP = \sum_{i\in|\U|} \delta^{\fsp}_i / |\U|$. As stated in (\ref{eq:fsp_1}), a precondition for satisfaction is being connected. In other words, $u_i$ must have a signal from BS$_j$ with a strength that is higher than $\snir^{\textrm{min}}$ and the perceived throughput must be at least $\rate^{\textrm{min}}_i$ bps. 

\section{Dutch cellular networks as a case study}\label{sec:perfeval}
In this section, we first introduce the datasets and model parameters we use to simulate the considered cellular networks. 
Afterwards, we provide an analysis of the current state of the network in terms of FDP and FSP performance, and show characteristics of the three main operators in the Netherlands.

\subsection{Datasets and simulation setting}
\label{sec:data}

\begin{figure}
    \centering
    \includegraphics[width = 0.42\textwidth]{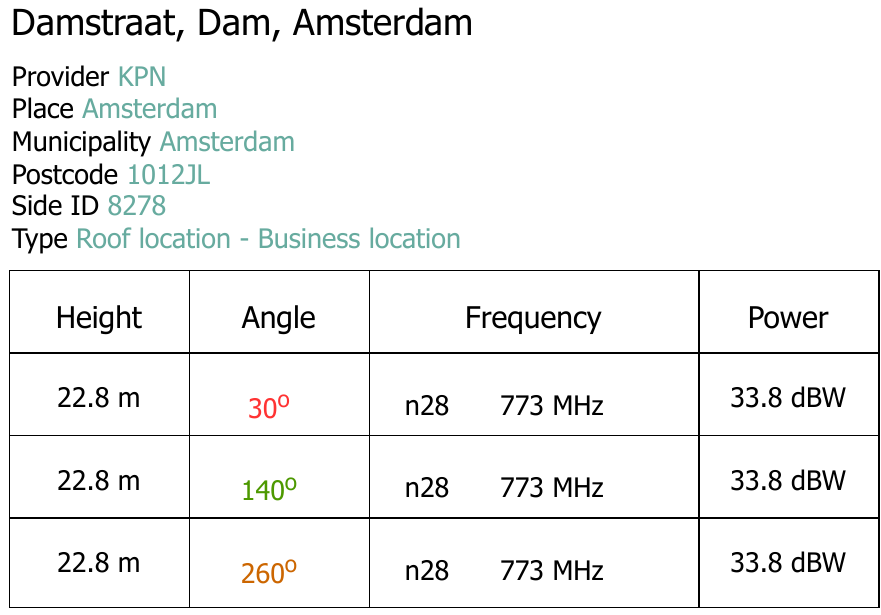}
     \caption{An example entry from the data set. For every BS, the following information is available: the  provider, place, height, angle, frequency,  power, \newInfo{reproduced from}~\cite{antennekaart}. }
    \label{fig:BS_example}
\end{figure}

\textbf{Antenna dataset:} We use the Dutch Telecommunication Authority's~({\color{black}Rijksinspectie Digitale Infrastructuur}) antenna registration dataset~\cite{antenneregister}, which includes the following information for each BS registered in the dataset: technology~(2G, 3G, 4G, 5G), location, center frequency, effective radiated power~(ERP) per channel, antenna height and antenna sectors. 
As Fig.~\ref{fig:BS_example} shows, BSs typically operate using three sector antennas, whose antenna gain model is introduced in Sec.~\ref{sec:system_model}.
%
To calculate the gain of these antennas, we use the antenna radiation pattern as given in the 3GPP TR 36.942 specification~\cite{access2020radio}. The main direction in which the antennas transmit is available in the data set. 
Since the dataset includes the effective isotropic radiated power (EIRP), we assume that this power already includes $G_{\text{max}}$ in \eqref{eq:AH}. 
Thus, to obtain $G_{ij}$, we only add \eqref{eq:AH} from the given EIRP.
The horizontal boresight angle of all antennas is given in the dataset.
We remove the omnidirectional antennas in the dataset, as manual investigation showed that these antennas are mostly placed indoors, and some MNOs choose to not register these antennas in the dataset \cite{antennekaart}. These removed data correspond to $5.3\%$ of the 3G, 4G and 5G BSs.
Moreover, we have removed BSs with 2G technology~(corresponding to $29\%$ of all BSs) from the dataset as 2G network serves a different purpose, e.g., such as smart metering, rather than voice or data communications. Moreover, 2G networks will be phased-out in the near future in the Netherlands~\cite{2gphasedout}.  

\textbf{Mobile network operators:}
\newInfo{The Netherlands has three MNOs: KPN, Vodafone, and Odido~(previously, T-Mobile). Additionally, many virtual MNOs use the infrastructure of these three operators. 
As the antenna data set does not provide data on the owner of each BS, we map each BS to one of the three MNOs based on the frequency range that each MNO owns \cite{frequentieoverzicht} (Table \ref{tab:frequency_statistics}). Our dataset includes information on the three MNOs which we refer to in this study as \KPN, \TMobile, and \Vodafone, in no particular order. Table \ref{tab:frequency_statistics} also shows the number of BSs of these three MNOs.  }

As can be seen from the operation frequencies listed in Table~\ref{tab:frequency_statistics}, at the time of this study, 5G deployments operate only in the low-band ($<$ 1 GHz) and mid-band 5G spectrum, i.e., 1 GHz to 2.6 GHz. Higher frequencies, such as 3.5 GHz, are expected to become available after an auction in December 2023\footnote{\url{https://www.rijksoverheid.nl/actueel/nieuws/2022/05/12/adviescommissie-35-ghz-band-in-2023-in-gebruik-voor-mobiele-communicatie}}. 
We believe that our analysis can also provide insights to the MNOs on where the higher frequency cells should be deployed for capacity and coverage improvement. 

\begin{table*}[]
\centering
\caption{Frequency bands and number of BSs per provider for 3G, 4G and 5G technologies.}
\label{tab:frequency_statistics}
\renewcommand{\arraystretch}{1.2}
\resizebox{1.8\columnwidth}{!}{%
\begin{tabular}{|l|lllc|llll|}
\hline
 &
  \multicolumn{4}{c|}{Centre frequency and bandwidth in MHz} &  \multicolumn{4}{c|}{Number of BSs}\\
 &
  \multicolumn{1}{c}{3G} &
  \multicolumn{1}{c}{4G} &
  \multicolumn{1}{c}{5G} & 
  \multicolumn{1}{c|}{Total (MHz)} & 
  \multicolumn{1}{c}{3G} &
  \multicolumn{1}{c}{4G} &
  \multicolumn{1}{c}{5G} &
  \multicolumn{1}{c|}{Total}   \\ \hline
\KPN &
942.2 (5), 2152.6 (5) &
  \begin{tabular}[c]{@{}l@{}}816 (10), 1474.5 (15), 1815 (20), \\ 2160 (20), 2605 (30), 2660 (10)\end{tabular} &
  \begin{tabular}[c]{@{}l@{}}773 (10), \\ 2160 (20)
  \end{tabular}  & 175 &4716 &4621 &3508 &12845 \\ \hline 
\TMobile &
 957.4 (5),  2137.4 (10) &
  \begin{tabular}[c]{@{}l@{}}796 (10), 950 (10), 1487 (10), \\ 1850 (10), 1860 (30), 1865 (20), \\ 2137.5 (15), 2580 (20), 2652 (4),\\ 2572.5 (15), 2672.5 (15), 2675 (20)\end{tabular} &
  783 (10) & 204   & 4752      & 4855   & 3590    & 13197   \\ \hline 
\Vodafone & \quad\quad\quad\quad---
  &
  \begin{tabular}[c]{@{}l@{}}763 (10), 806 (10), 1459.5 (15), \\ 1835 (20), 2117.5 (15), 2120 (20), \\ 2630 (20), 2644.4 (10)\end{tabular} &
  1835 (20)& 140  & 0   & 4477   & 3395   & 8313 \\ \hline
\end{tabular}%
}
\end{table*}

\textbf{BS transmission power:} The power provided in the dataset is the maximal EIRP that the BS can transmit~(in dBW), which is an upper bound on the EIRP that will actually be used. 
Comparing the values with the reported values in \cite{deruyck2014power} which are collected from data sheets of 
network equipment manufacturers, we infer that the power values in the dataset represent the total power budget of a BS which is used for also other tasks such as air cooling or digital signal processing. Hence, we assume that every BS operates using $90\%$ of its maximum power recorded in the dataset. {We have observed no significant differences in the power levels of different MNOs.}

\textbf{Rate requirements:}
In a cellular network, users can have different rate requirements based on their application. 
To represent the rate requirements, we assume a varying rate requirement, where each user gets assigned a rate requirement $C^\text{min}$ that is uniformly distributed in [$R_{\text{min}}$, $R_{\text{max}}$] Mbps.

\textbf{Population density for each $\mathbf{500\times500}$m square:}
To simulate the population, we use the data from Statistics Netherlands~(Centraal Bureau voor Statistiek)\cite{cbs}, which records the number of inhabitants per $500\times500$m square in the Netherlands and the \textit{urbanity} of this area {\color{black} at the end of 2020}. 

The urbanity data distinguishes between five levels of urbanity, where level-1 represents the highest address density and level-5 the lowest. More precisely, the \textit{area address density} quantifies the number of addresses within a circle with a radius of one kilometer around an address divided by the area of the circle\footnote{\url{https://www.cbs.nl/nl-nl/onze-diensten/methoden/begrippen/stedelijkheid--van-een-gebied--}.},
and distinguishes the following intervals for the urbanity levels: 
\begin{itemize}
  \item level-$1$ above 2500 per km$^2$;
  \item level-$2$ in [1500, 2500) km$^2$; 
  \item level-$3$ in [1000, 1500) per km$^2$; 
   \item level-$4$ in [500, 1000) km$^2$; 
\item level-$5$ lower than 500 per km$^2$.
\end{itemize}

Based on the urbanity level of the area in which a BS is located, we determine whether it is an urban macrocell~(UMa) or a rural macrocell~(RMa). 
We assume that an urbanity level of $1\,-3$ corresponds to the UMa scenario and levels 4 and 5 are rural areas (RMa). 

We assume that a fixed fraction $f_p$ of the population is active at a time and therefore $1-f_p$ fraction of the population uses other connectivity modes such as Wi-Fi or is not connected to a mobile network.  We simulate users in these $500 \times 500$m squares by a homogeneous Poisson point process with a fraction $f_p$ of the population density as intensity measure. As the number of customers for each MNO is not publicly available, we equally divide the users among the MNOs.\footnote{Based on Tables~\ref{tab:frequency_statistics} and \ref{tab:municipality_province_statistics}, one might conclude that an equal distribution among MNOs would not be fair. We also have studied a different user distribution: $40\%/40\%/20\%$ for MNOs 1, 2 and 3 and for this distribution the trend in the results stayed the same.}

\begin{table}[t]
\centering
\renewcommand\arraystretch{1.2}
\caption{Default parameters used in simulations.}
\label{tab:sim-params}
\begin{tabular}{|l|l|} \hline 
Parameter& Value(s)\\ \hline
Rate requirement $R_{\text{min}} - R_{\text{max}}$& 8 - 20 Mbps \\ \hline
Minimum SINR $\snir^\text{min}$ & 5 dB \\ \hline
Fraction of active population $f_p$ & $2\%$\\ \hline
Maximum distance for interference $r_\text{max}$ & $5$ km\\ \hline
\end{tabular}
\end{table}

\begin{figure*}[thb]
    \centering
   \subcaptionbox{Number of BSs per km$^2$.\label{sfig:bs_per_area_provinces}}{\includegraphics[width=.45\textwidth]{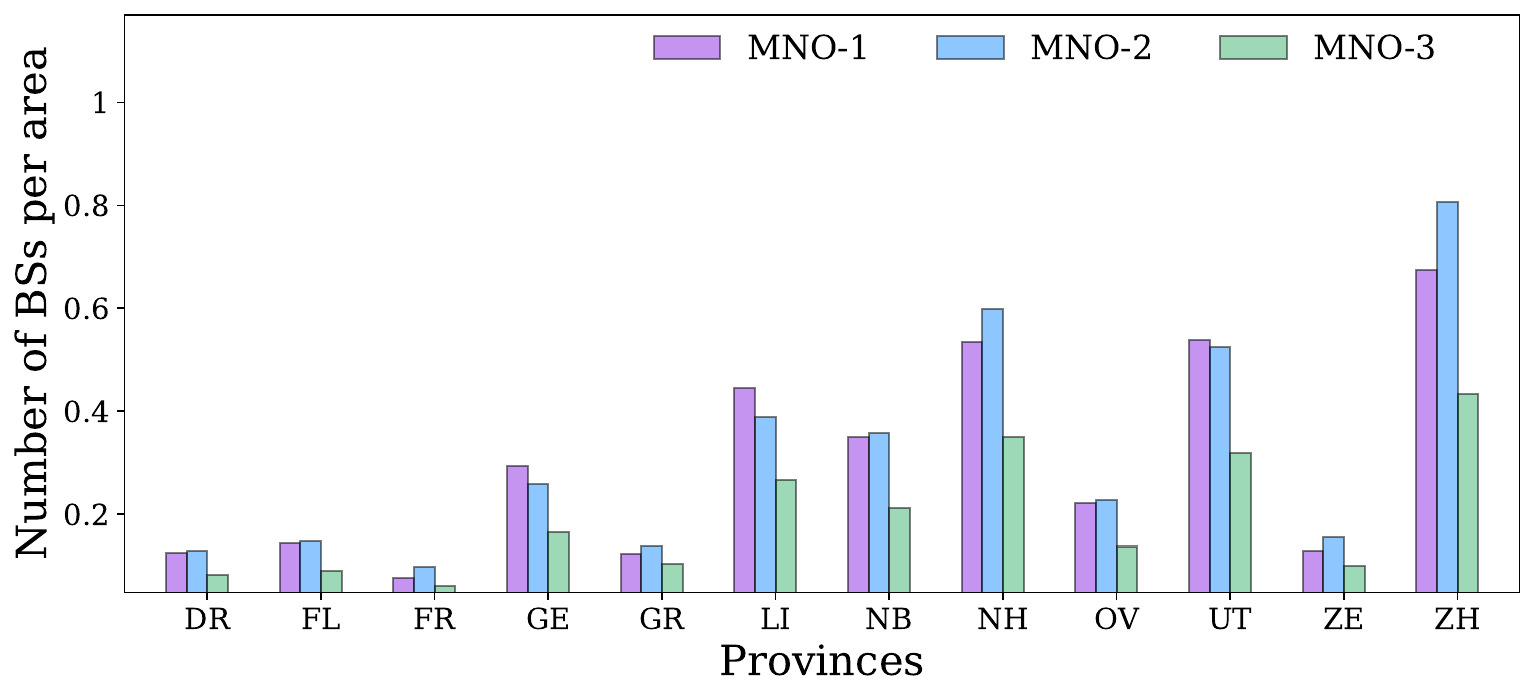}\hfill}\label{fig:statistics_regions}
    \subcaptionbox{Number of active users per BS.\label{sfig:user_per_BS_provinces}}{\includegraphics[width=.45\textwidth]{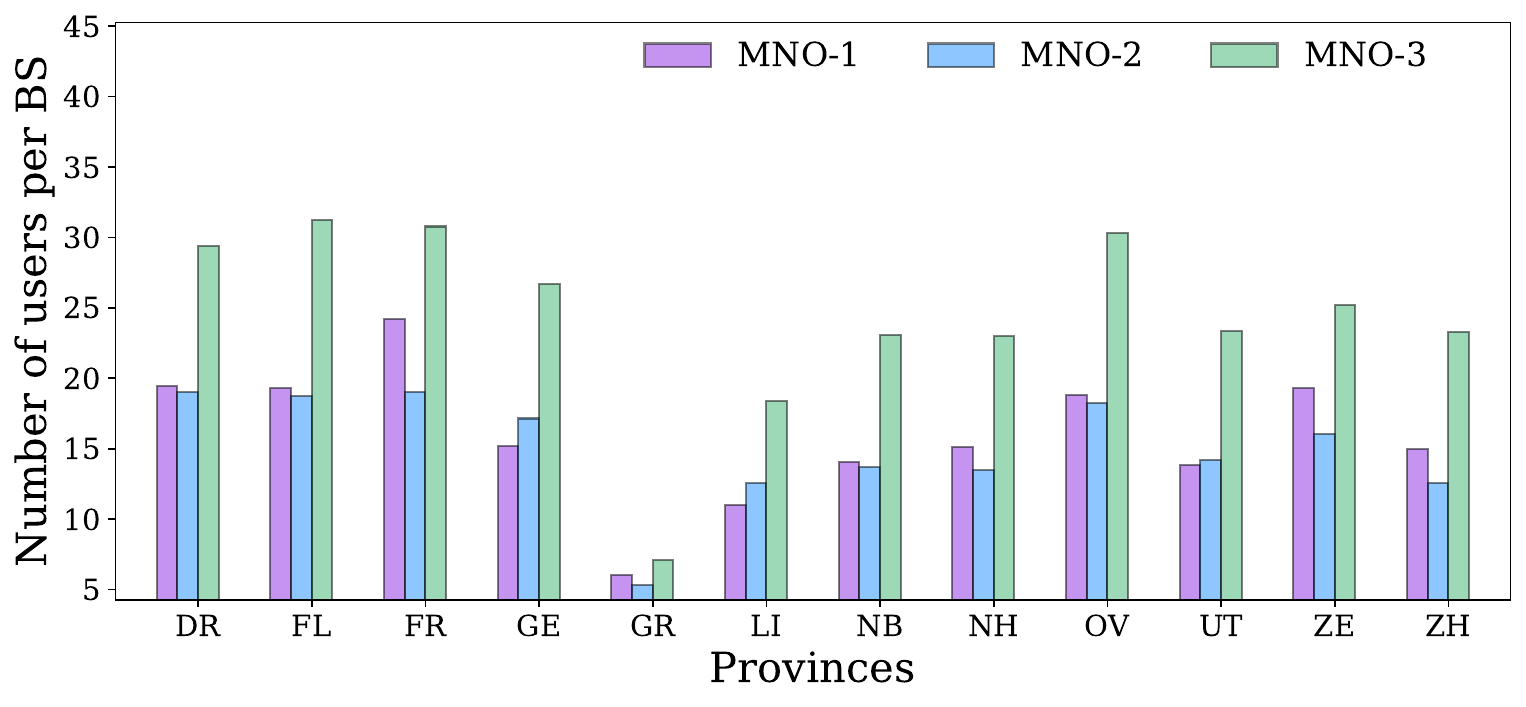}}
    \caption{Number of users per BS and BS density (km$^2$) per province.} 
\end{figure*}

\textbf{Failures:} To measure the resilience of the network, we simulate two types of failures: \textit{isolated} and \textit{correlated} failures. 
Technology-related failures or cyber attacks on the network can be modeled as \textit{isolated} failures, where every BS fails independently and with probability $p_{iso}$. 
Disasters such as earthquakes and hurricanes often only impact a specific geographic area. Therefore, they affect the performance of several nearby network links.
For this reason, we model this as \textit{correlated failures} where link within and close by the disaster area fail. 
We pick the center of the simulated area as center of impact, and let all BSs that are within radius $r$ meter of this center fail~\cite{agarwal2013resilience}.

\subsection{Current state of the networks} \label{subsec:perf-current}
Using the described datasets, we first investigate each MNO's performance separately without any failures using the FSP and FDP metrics. Next to that, we show some statistics about the current deployments, e.g., the number of users per BS or the number of BSs per km$^2$. 
We also consider a scenario with national roaming in which all networks work together to serve the users and each user can connect to the BS which provides the best performance according to (\ref{eq:user_association}).

Table~\ref{tab:sim-params} provides the default values of the parameters used in the simulations. In our evaluation, \newInfo{ we set $f_p=2\%$ to keep the average number of users per BS around $10$ \cite{george2018distribution}.}
Moreover, 
we consider $R_{\text{min}}=8$\,Mbps as the Dutch regulatory body asserts that each MNO must provide at least 8 Mbps outdoor data rate in each region~\cite{min8mbps}. \newInfo{As $R_{\text{max}}$, we choose $20$ Mbps considering the recommended application rate requirement in the literature, e.g., average 20 Mbps for mobile gaming\footnote{NordVPN \url{https://nordvpn.com/blog/internet-speed-for-gaming/}}.} 
Then, when users are associated to a BS, the BS calculates the effective bandwidth $\effectiveRes_{ij}$ per user according to \eqref{eq:effective_bandwidth}. 

\begin{table}[t]
\caption{Statistics of the 12 Dutch provinces.
\label{tab:municipality_province_statistics}}
\resizebox{\linewidth}{!}{%

\begin{tabular}{|l|ccc|c|c|c|}\hline 
   & \multicolumn{3}{c|}{Number of BSs}                                       & \multirow{2}{*}{\#users} & \multirow{2}{*}{Avg.urb.} & \multirow{2}{*}{Area, km$^2$}\\
   & \KPN & \TMobile & \Vodafone &                          &                           &      \\ \hline

DR & 337  & 345 & 223 & 6558  & 4.7 & 2691  \\ 
FL & 212  & 218 & 131 & 4089  & 3.5 & 1471  \\ 
FR & 272  & 346 & 214 & 6586  & 4.6 & 3555  \\ 
GE & 1508  & 1334 & 856 & 22877  & 4.0 & 5145  \\ 
GR & 291  & 330 & 247 & 1761  & 4.4 & 2378  \\ 
LI & 986  & 863 & 590 & 10859  & 4.0 & 2214  \\ 
NB & 1773  & 1818 & 1081 & 24947  & 3.7 & 5076  \\ 
NH & 1538  & 1723 & 1008 & 23215  & 3.1 & 2875  \\ 
OV & 760  & 783 & 471 & 14292  & 4.0 & 3427  \\ 
UT & 841  & 820 & 498 & 11646  & 3.1 & 1562  \\ 
ZE & 244  & 294 & 187 & 4716  & 4.5 & 1889  \\ 
ZH & 1931  & 2311 & 1244 & 28991  & 2.5 & 2865  \\ \hline
\end{tabular}%
}
\end{table}

{Each simulation is repeated 100 times for statistical significance as user locations and path loss are random.} We provide simulations of the performance of national roaming on the province level.  Table~\ref{tab:municipality_province_statistics} provides an overview of different properties of all provinces. 
Fig.~\ref{sfig:bs_per_area_provinces} shows the BS density per km$^2$ in each province. We can observe clear differences across provinces and among MNOs. 
First, \Vodafone~has the lowest BS density due to its significantly lower number of BSs as highlighted in Table~\ref{tab:municipality_province_statistics}. As expected, ZH as the most populated province~(including Amsterdam) has the highest BS density. Comparing \KPN and \TMobile, we do not observe a significant difference.  Fig.~\ref{sfig:user_per_BS_provinces} shows the higher user density per BS for \Vodafone in all provinces compared to \KPN and \TMobile due to its less dense BS deployment.

\section{Performance analysis} 

\begin{figure*}[thb]
    \centering
    \subcaptionbox{FDP\label{sfig:FDP}}{\includegraphics[width=.33\textwidth]{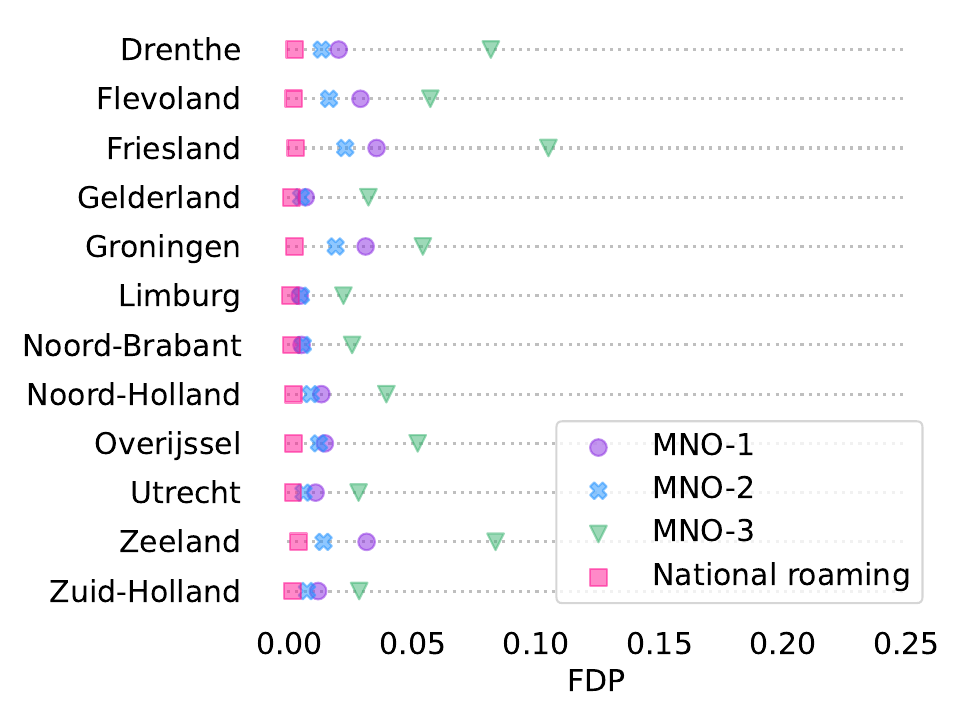}}\hfill
     \subcaptionbox{FSP\label{sfig:FSP}} {\includegraphics[width=.33\textwidth]{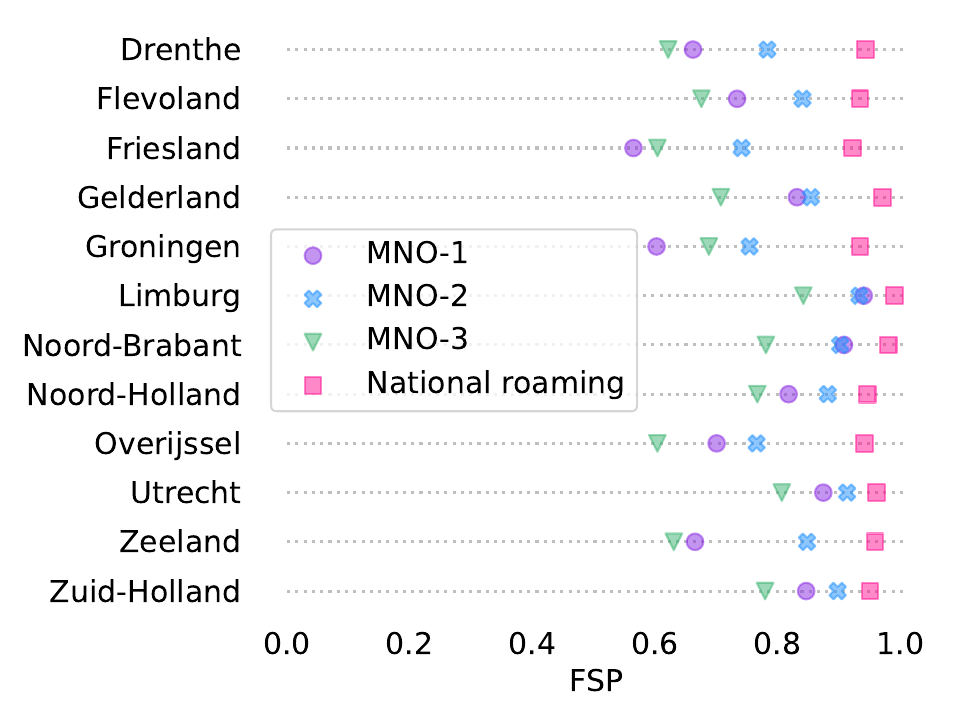}}\hfill
    \subcaptionbox{Gains of each MNO from national roaming.\label{sfig:FDP_gains}}{\includegraphics[width=.33\textwidth]{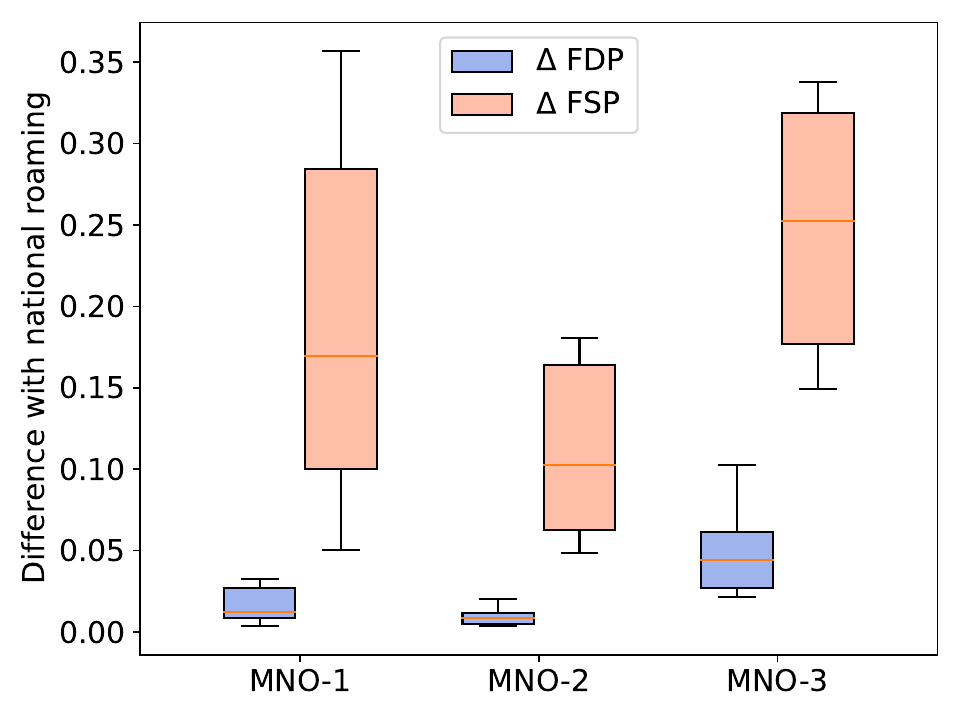}}\hfill
    \caption{FDP, FSP, and achieved gains from national roaming for the three MNOs in different provinces.} %
    \label{fig:FDPFSP}
\end{figure*}

With these insights on the infrastructure of each MNO, now, let us discuss the coverage and capacity performance of these MNOs. Fig.~\ref{fig:FDPFSP} shows the FSP and FDP per province per MNO and for all MNOs together, representing the case where all operators can use each other's network as in national roaming. 
Looking closer at Fig.~\ref{sfig:FDP}, we have the following three observations. \textit{First}, in various regions, some operators fail to provide sufficient SINR to their customers, resulting in an FDP as high as 0.11 in Friesland or around 0.09 
in Zeeland. 
Especially in these regions, national roaming provides its benefits as reflected in a significant improvement in FDP with an achieved FDP of zero. Moreover, \TMobile consistently achieves a higher performance in almost all regions compared to \KPN, whose performance is in turn significantly better than \Vodafone. 
We attribute this superior performance of \TMobile to its higher spectrum resources~(204 MHz vs. 175\,MHz and 140\,MHz) as we have not observed a significant difference in their BS density in Fig.~\ref{sfig:bs_per_area_provinces}. The performance gap between \Vodafone and other MNOs can emerge due to the lower spectrum resources and lower BS density. 

When it comes to FSP, Fig.~\ref{sfig:FSP} again shows a superior performance of \TMobile. However, the achieved FSP varies between $0.73$ and $0.89$, indicating a need for performance improvement. Comparing \KPN and \Vodafone, generally speaking, \KPN outperforms \Vodafone in terms of FSP, except in Friesland, which is one of the worst-performing regions in terms of both FSP and FDP. 
These low-FSP regions could be considered as initial places for investment to ensure higher user satisfaction. Also, \KPN and \Vodafone provide a more varying FSP across provinces compared to \TMobile. For example, the achieved FSP for \TMobile ranges from $0.70 - 0.95$, while for \KPN and \Vodafone, the achieved FSP lies in $[0.54, 0.89]$ and $[0.56,0.80]$, respectively. This large range in satisfaction might emerge as a result of different deployment strategies.

\textit{Second}, letting users connect to every BS regardless of their MNO improves FSP consistently as expected. This improvement is, however, more significant in some regions such as Friesland or Overijssel. %
\textit{Third}, national roaming improves FSP by around $0.05-0.35$ in absolute terms~(Fig.~\ref{sfig:FDP_gains}). However, in contrast to FDP, it does not yet suffice to meet all rate requirements as reflected by FSP always being below $1.0$. This can be considered as an indication of the need for infrastructure expansion or for more advanced schemes to provide higher throughput, e.g., expanding to higher spectrum bands with abundant bandwidth. 
When it comes to gains experienced by each MNO in the case of national roaming, Fig.~\ref{sfig:FDP_gains} plots the performance gain in terms of FDP and FSP observed by each MNO. In line with our earlier observations, \Vodafone benefits the most from national roaming, followed by \KPN with a slight difference over \TMobile. Please note that despite $\Delta$FDP being very narrow for \KPN and \TMobile, the resulting FSP gain is still remarkable also for these operators.
Note that all $\Delta$FSP and $\Delta$FDP values in Fig.~\ref{fig:FDPFSP} are positive implying that national roaming does not lead to performance degradation  and even the best-performing MNO can benefit from it, albeit less significantly compared to other MNOs with less-dense deployment. 
{As a nation-wide implementation of national roaming requiring collaboration of all MNOs and technologies might be hard to realize, we also investigated roaming in a limited way to provide insights to the MNOs and  when and where they could benefit most from national roaming. Such an analysis could also help national telecommunication regulatory bodies to develop policies enforcing national roaming for serving the underserved areas or regions benefiting the most from this mode of operation. We focused on the following questions}\footnote{In a recent work, we also compared national roaming as considered in this paper with current implementation of national roaming where MNOs only can use others' network when they experience a failure in their own network, i.e., as a fallback strategy. Please refer to \cite{weedage2023national} for more details.}: {(i) which cellular technology, i.e., 3G, 4G or 5G, should be prioritized for roaming if MNOs are eager to develop such roaming agreements? and (ii) in which areas should the MNOs consider roaming agreements, i.e., rural areas where the infrastructure is less dense or in urban areas where the population density thereby the traffic load is higher?  }

\begin{figure*}[thb]
    \centering
    \subcaptionbox{FDP\label{sfig:technology_fsp}}{\includegraphics[width=.45\textwidth]{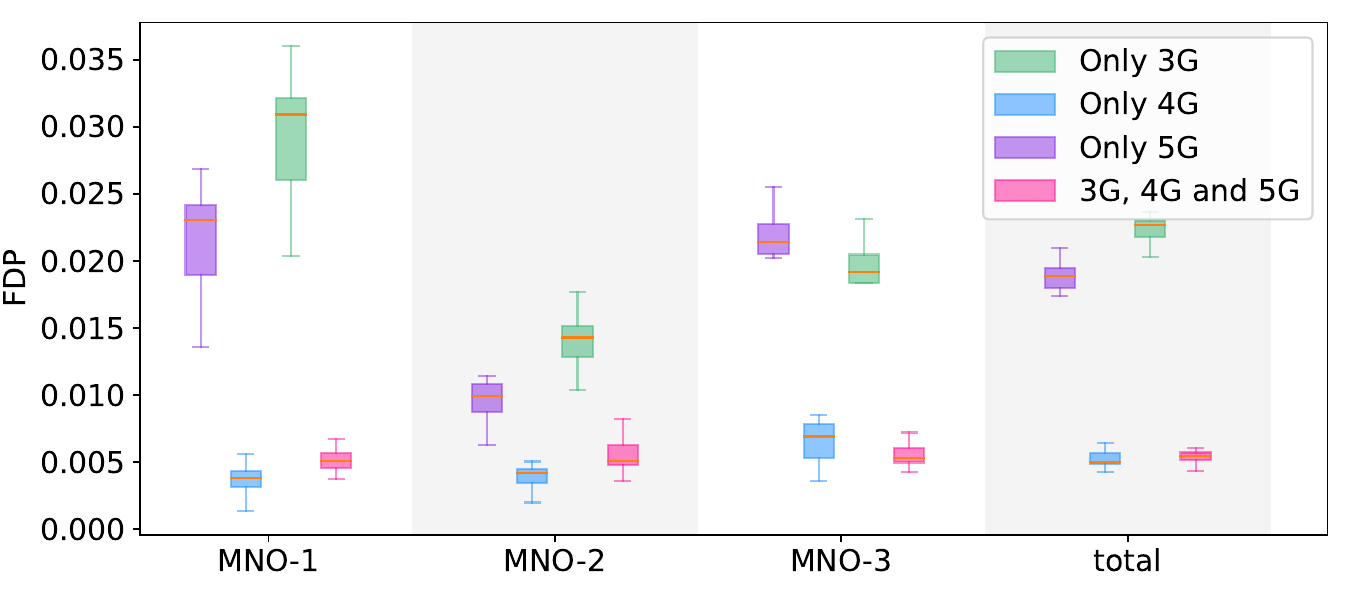}}\hfill
    \subcaptionbox{FSP \label{sfig:technology_fdp}}{\includegraphics[width=.45\textwidth]{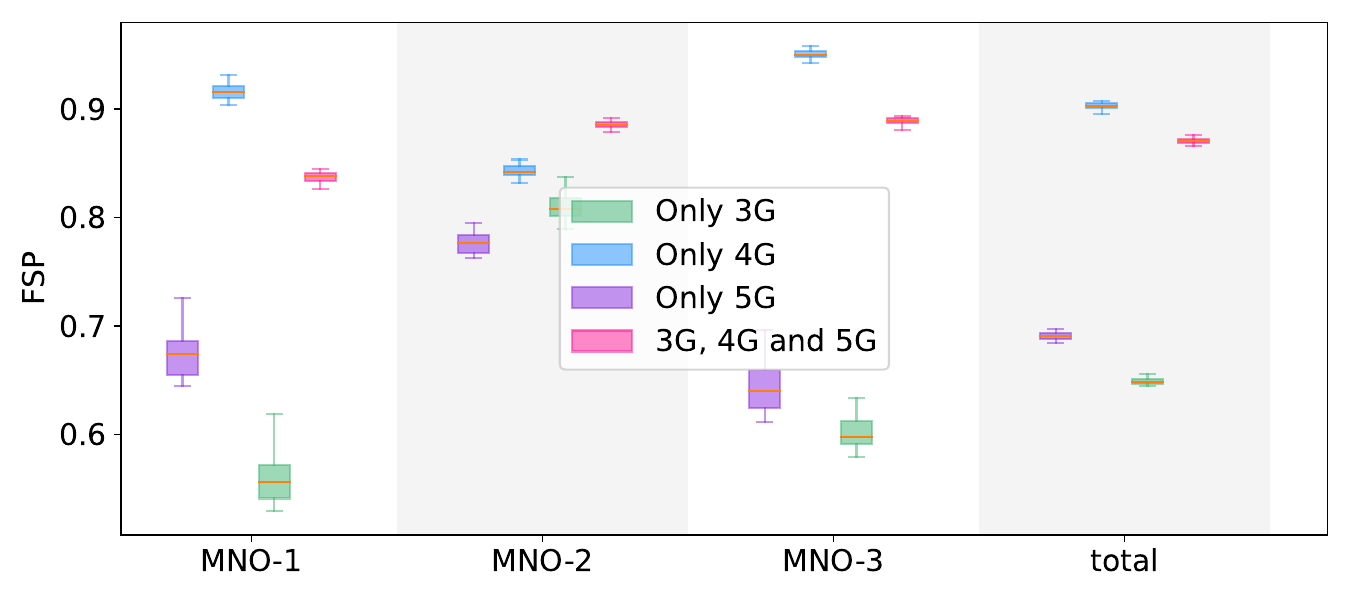}}\hfill
    \caption{FDP and FSP in Amsterdam under only roaming for a specific technology~(5G, 4G or 3G).  
    } 
    \label{fig:technology_Enschede}
\end{figure*}

{Fig.~\ref{fig:technology_Enschede} depicts the resulting FDP and FSP when operators only share BSs of a certain technology to serve each other's users. In this case, users access to the network of their own operator for the remaining technologies for which national roaming is not implemented. Moreover, we compare this to the scenario with sharing all technologies (`3G, 4G and 5G'). This figure shows that only sharing the 4G LTE technology performs similar to the full national roaming approach, and even slightly better in terms of FSP for \KPN and \Vodafone. Only sharing 5G NR or 3G UMTS performs significantly worse, both in terms of FDP and FSP. We speculate that the superiority of 4G LTE can be explained by the large spectrum resources (Table \ref{tab:frequency_statistics}), as this will result in the least interference. For \KPN, while 36\% of the cell sites are 4G LTE sites, the spectrum for these areas accounts for 60\% of the spectrum used by this operator. For \TMobile, 40\% of the cell sites are 4G operating on 88\% of the spectrum owned by \TMobile. Lastly, for \Vodafone, 4G cell sites correspond to 54\% of the cell sites and the spectrum allocated to these cells is 86\%. }

\begin{figure*}[thb]
    \centering
    \subcaptionbox{FDP\label{sfig:area_fsp}}{\includegraphics[width=.45\textwidth]{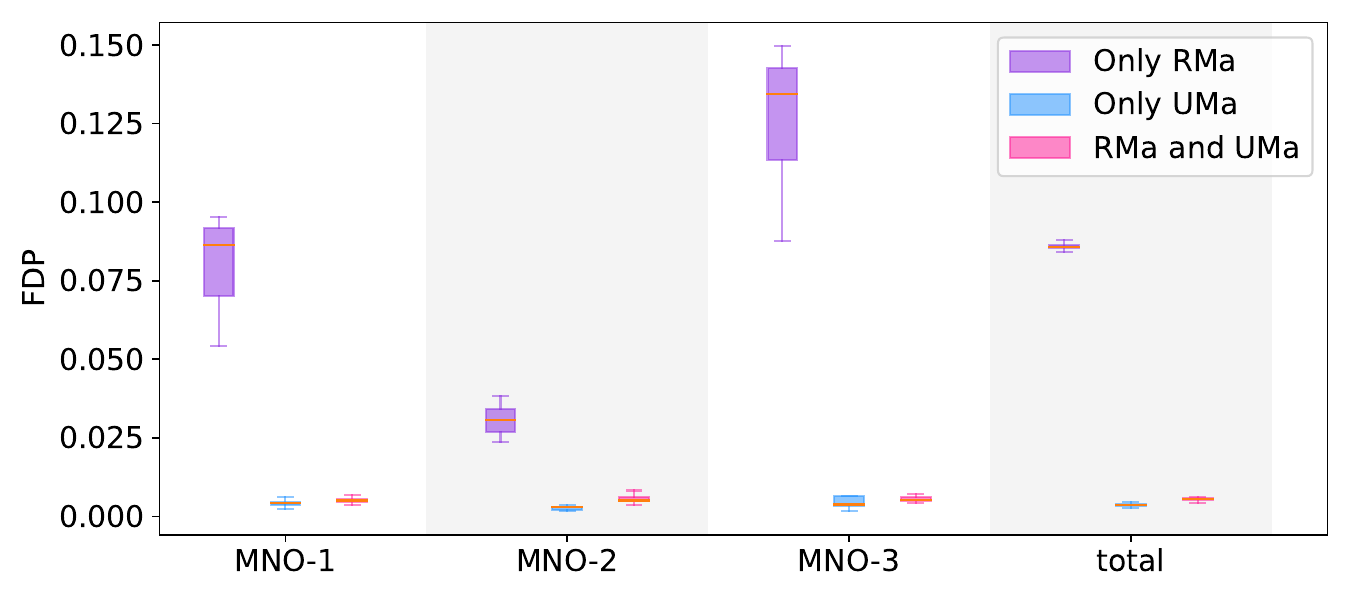}}\hfill
    \subcaptionbox{FSP \label{sfig:area_fdp}}{\includegraphics[width=.45\textwidth]{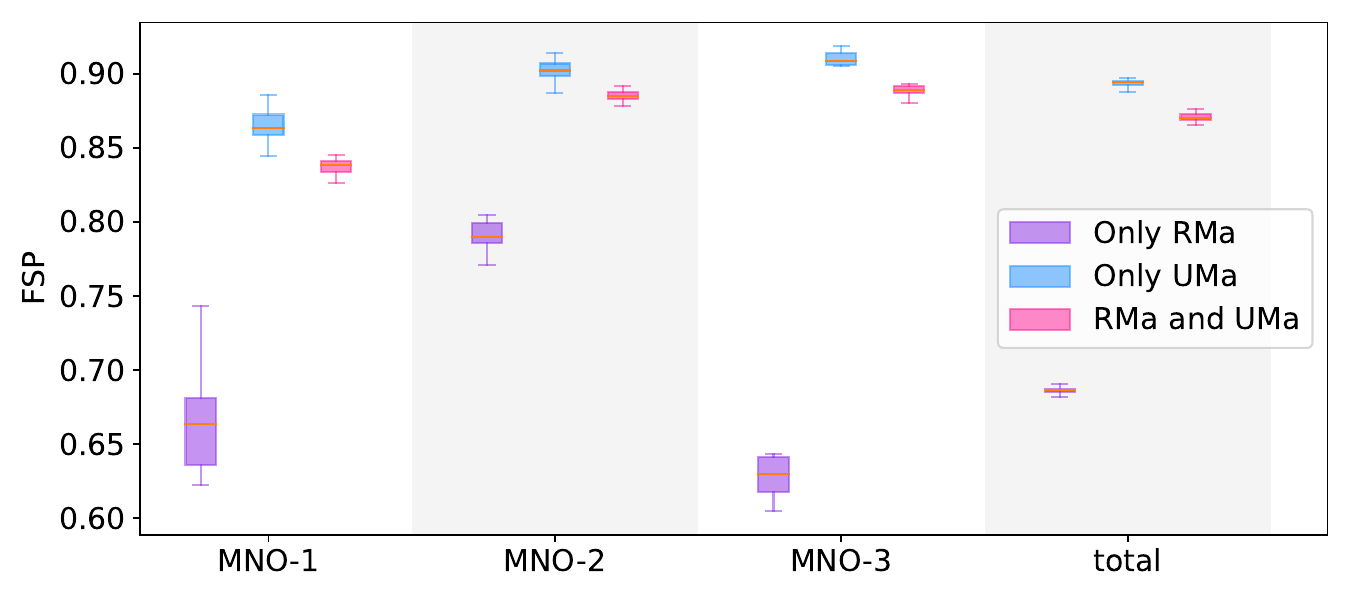}}\hfill
    \caption{FDP and FSP in Amsterdam for only sharing in specific cell types (UMa or RMa).} 
    \label{fig:area_Enschede}
\end{figure*}

{For addressing (ii), we considered roaming only in UMa or RMa cells as defined in \cite{3gpp38901} as these two cells correspond to urban areas and rural areas, respectively. Fig.~\ref{fig:area_Enschede} illustrates FDP and FSP when sharing is applied only in certain areas. This figure shows that sharing infrastructure in the UMa areas outperforms sharing in RMa areas significantly. We attribute this result to the number of resources: for all operators, there are 5-6 times more BSs classified as UMa compared to RMa. Moreover, the population is concentrated in urban areas resulting in higher traffic load. 
A clear recommendation on whether MNOs should prioritize sharing a certain technology or sharing in certain areas is not straightforward as our results suggest that sharing more resources (either spectrum or BS locations) are expected to result in better FDP and FSP. }

\begin{figure}
    \centering
    \includegraphics[width = 0.3\textwidth]{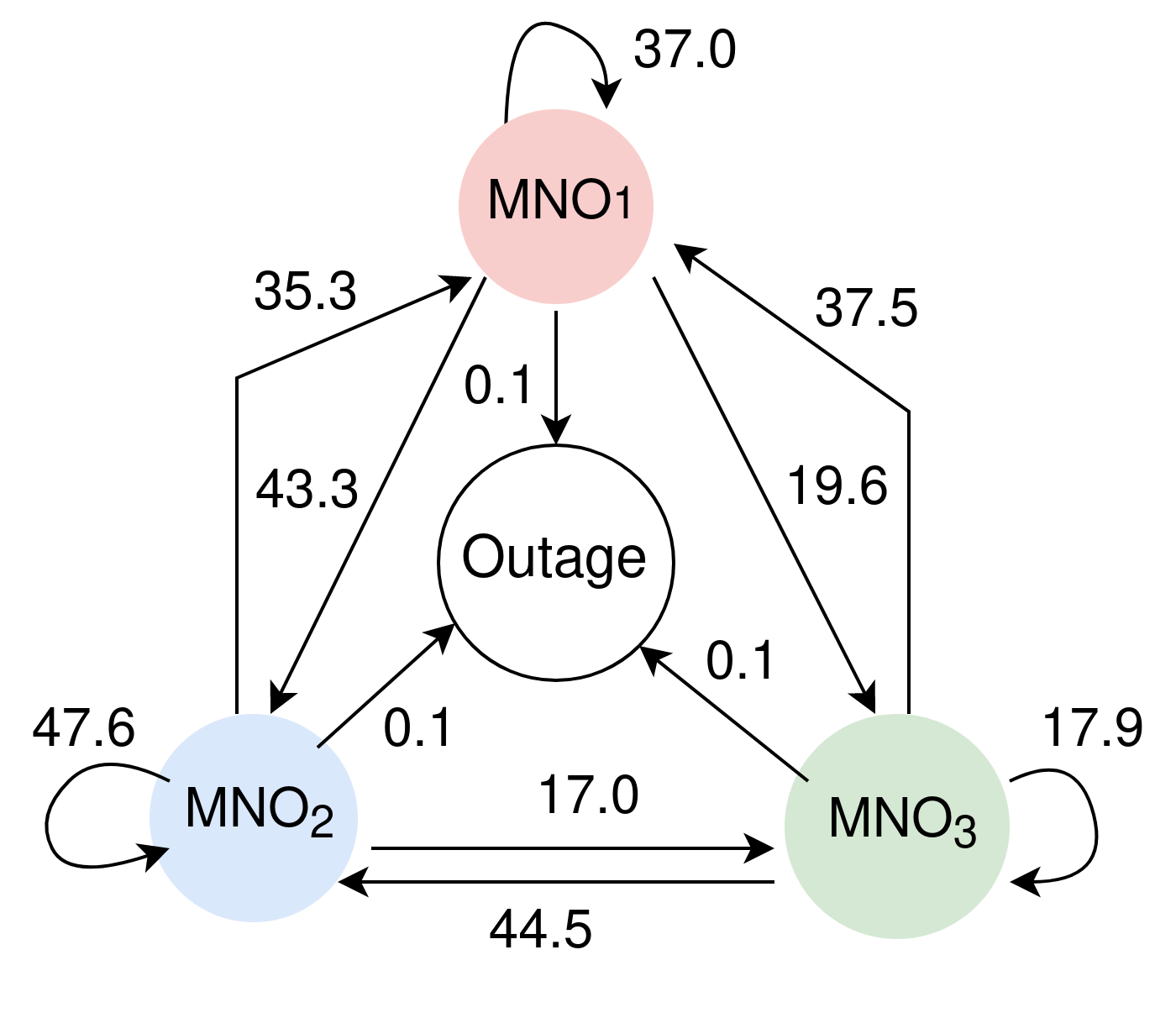}
    \caption{Percentage of users in Enschede being served
by each MNO. An arrow shows the percentage of users
of an MNO being roamed to another MNO.}
    \label{fig:NR-percentages}
\end{figure}

\textbf{User association: }
{Fig.~\ref{fig:NR-percentages} shows the percentage of users that is being roamed to another MNO in case of national roaming. This figure shows that under national roaming,  user association changes significantly and at most half of the users of an MNO will connect to that MNO, while the other users connect to the BSs of another MNO. For \Vodafone, this percentage is even lower: only 17.9 $\%$ of their users connect to the BSs of their own MNO. These results are in line with the results shown in Fig.~\ref{fig:FDPFSP}: since \Vodafone provides less coverage to their users, they benefit most from national roaming and therefore many users will use the network of \KPN and \TMobile.}

\begin{tcolorbox}[title=\textbf{Takeaway},colback=gray!20,colframe=white,arc=3mm,boxsep=3mm]
 Our analysis shows that FDP and FSP vary across MNOs and geographic regions. National roaming can consistently offer benefits; up to 13\% improvement in FDP and up to 55\% in FSP. However, the observed benefits vary across MNOs, technologies and regions. Thus, for internet equity, {national roaming can be a solution in those regions not meeting a desired level of FDP and FSP to achieve a certain target level.
Moreover, the simulations show that only sharing a certain technology or in a certain area results in better FDP and FSP, although amount of resources (available spectrum or number of BSs) in this case is the most important factor: sharing more resources generally results in better FDP and FSP.} 
Similarly, MNOs can consider these areas for their own network expansion with newer technologies, including mid-band 5G frequency usage for enhancing FDP and high-band 5G frequencies for enhancing FSP. 
\end{tcolorbox}

\section{Resilience under failures} \label{sec:perf-failures}
Now, we investigate how failures might affect each MNO in terms of FDP and FSP and how national roaming could help in these cases.  
In case of an isolated failure, a cell tower might fail due to software errors~(e.g., misconfiguration or malicious attacks) or hardware errors~(e.g., power loss) independent of the other towers. Also, this type of failure represents the case where MNOs conduct regular maintenance on their network, during which some BSs become out-of-service. 
Second, correlated regional failures represent failures in a spatial locality due to certain events, e.g., a thunderstorm in a smaller region or an earthquake or flood affecting a larger region. Failures on the backhaul transport network can also be considered in this category, as such failures affect multiple BSs in a certain region simultaneously~\cite{yanPredictiveImpact,vass2021probabilistic}. In this case, BSs located in the same region will be affected similarly. 
For the isolated failures, we test a scenario in which a fraction $p_{\text{iso}}$ of the BSs fails. In the case of a correlated regional failure, all BSs within a circle of radius $r_{\text{fail}}$ meters of the center fail, where we assume the center is the centroid of the region. 

We conduct simulations for these scenarios on the municipality level, each with 100 independent runs for statistical significance. As municipality, we choose Enschede since it is a middle-sized municipality with both urban and rural areas. 
However, the general results for Enschede are similar to every other municipality and can be found in  \cite{interface}.\footnote{\newInfo{The impact of failures for other cities and provinces can be found on our interactive resilience map of the Netherlands \cite{interface}.}}

\begin{figure*}[htb]
    \centering
    \subcaptionbox{FDP\label{sfig:random_failure_NR_FDP}}{\includegraphics[width=.45\textwidth]{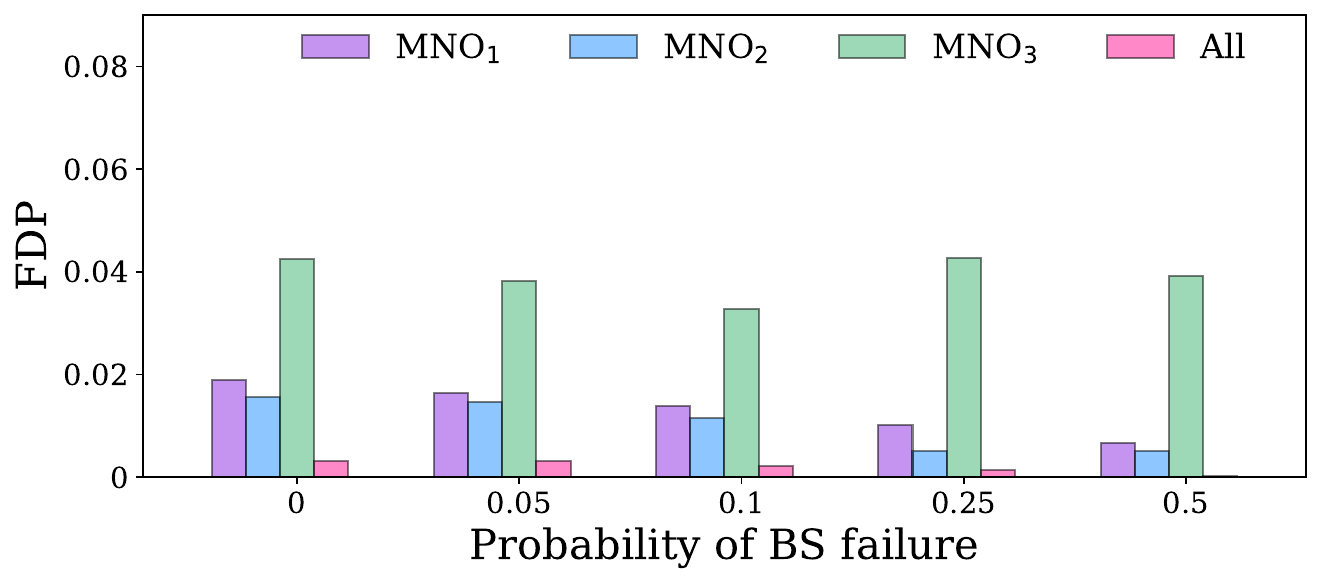}}\hfill
    \subcaptionbox{FSP \label{sfig:random_failure_NR_FSP}}{\includegraphics[width=.45\textwidth]{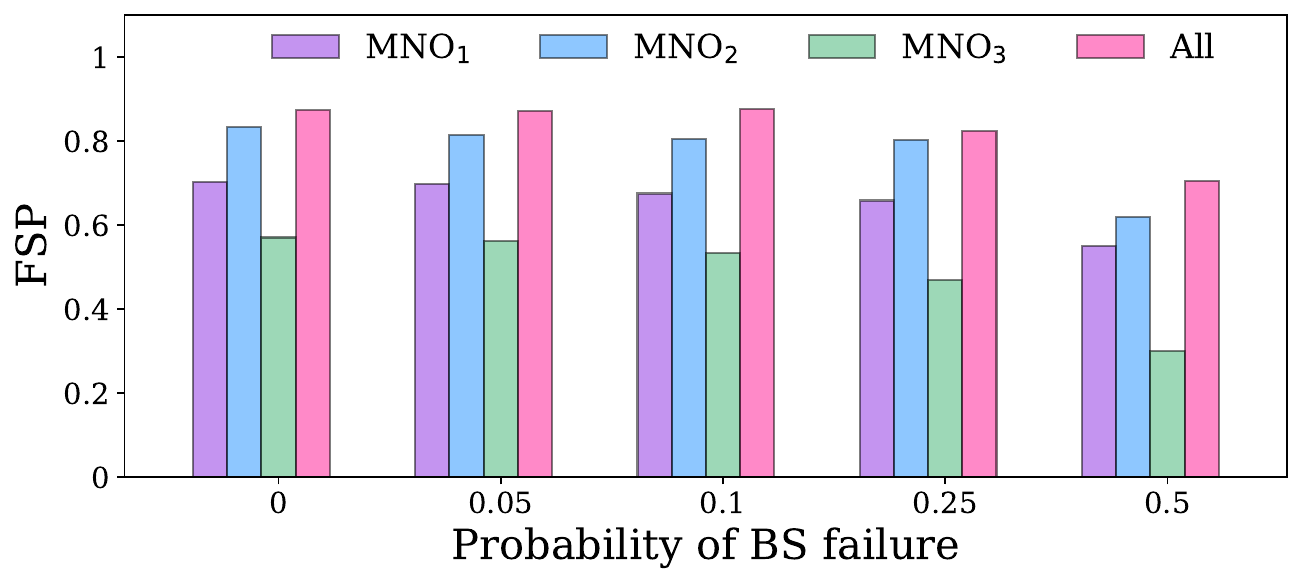}}\hfill
    \caption{FDP and FSP under isolated failures in Enschede for different MNOs, where each BS fails with probability $p_{\text{iso}}$.    \label{fig:random_failure_nationalroaming} }

\end{figure*}

\textbf{Isolated failures:} Fig.~\ref{fig:random_failure_nationalroaming} shows each MNO's performance in Enschede in terms of FDP and FSP under isolated failures, where every BS fails independently with probability $p_{\text{iso}}$. Note that $p_{\text{iso}}=0$ corresponds to a scenario without failures. Comparing the performance with this baseline scenario, 
as observed in earlier studies such as \cite{yanPredictiveImpact}, we infer that individual failures do not have a significant impact on the end users due to the inherent signal coverage redundancy in the network. 
However, contradicting the intuitions, Fig.~\ref{sfig:random_failure_NR_FDP} shows that higher $p_{\text{iso}}$ might result in lower FDP.
For instance,  for $p_{\text{iso}}=0.25$, FDP is lower for \KPN and \TMobile compared to the maintained FDP for $p_{\text{iso}}=0$.
A closer investigation shows that this is due to a decrease in the interference in the system with the decrease in the number of BSs. For \KPN, the difference between the received SINR and the SNR is $12$dB in a scenario without failures while it is only $9$dB for $p_{\text{iso}} = 0.25$. In other words, the interference decreases with failing BSs resulting in higher SINR on the average leading to lower FDP. However, interference management plays a key role in maintaining a high signal quality and consequently high capacity. Our observations are based on the assumption that MNOs implement interference management schemes and the closest three co-channel BSs do not interfere with each other. However, under other assumptions or a more advanced frequency re-use scheme, these results could be different.
When it comes to FSP, Fig.~\ref{sfig:random_failure_NR_FSP} suggests that users experience service quality degradation more drastically if MNOs do not implement infrastructure sharing. Failures up to $10\%$ of the BSs do not affect the networks significantly, but for higher values of $p_{\text{iso}}$ the surviving BSs become overloaded~(i.e., has to serve an increased number of users) which causes degradation in user satisfaction represented by lower FSP. 
Note that the decrease in FSP can be mitigated by dynamic frequency allocation schemes which re-allocate the frequency resources of the failing BSs to the active ones. 
Comparing the benefit of infrastructure sharing under normal operation~($p_{\text{iso}}=0$) against that of under failures~($p_{\text{iso}}>0$) by accounting for both FDP and FSP, we can conclude that sharing 
leads to better performance under failures. The networks of \KPN and \TMobile in particular are sufficiently redundant to still ensure coverage under failures. However, to also ensure satisfaction, network sharing is paramount.

\begin{figure*}[htb]
    \centering
    \subcaptionbox{FDP\label{sfig:correlated_failure_NR_FDP}}{\includegraphics[width=.45\textwidth]{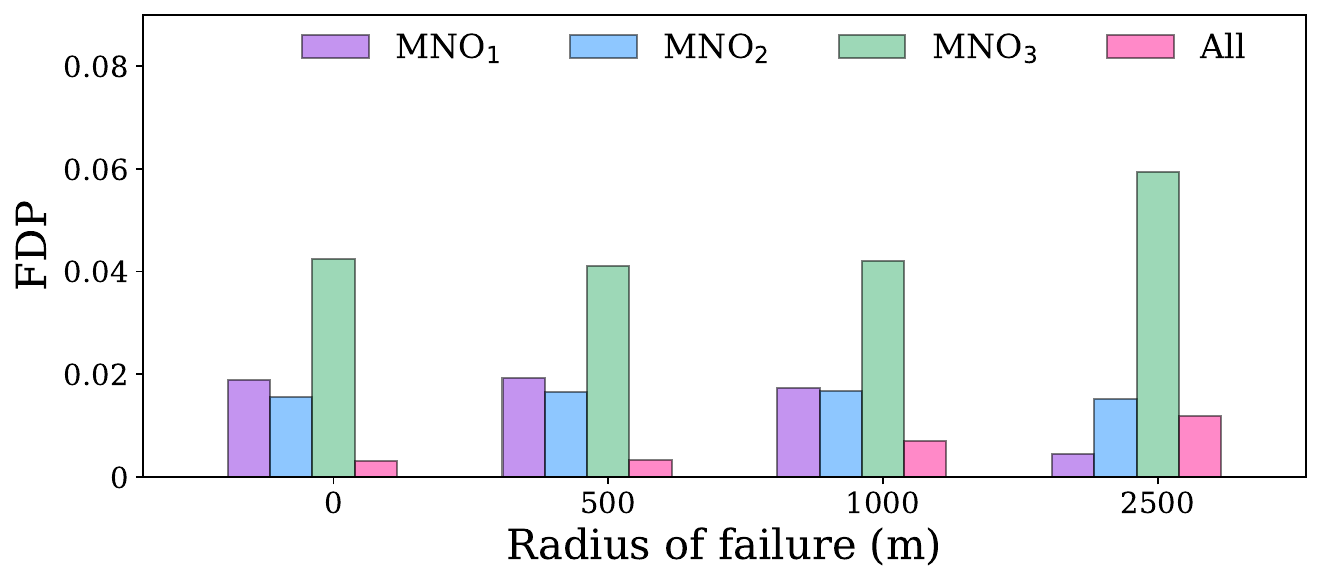}}\hfill
    \subcaptionbox{FSP \label{sfig:correlated_failure_NR_FSP}}{\includegraphics[width=.45\textwidth]{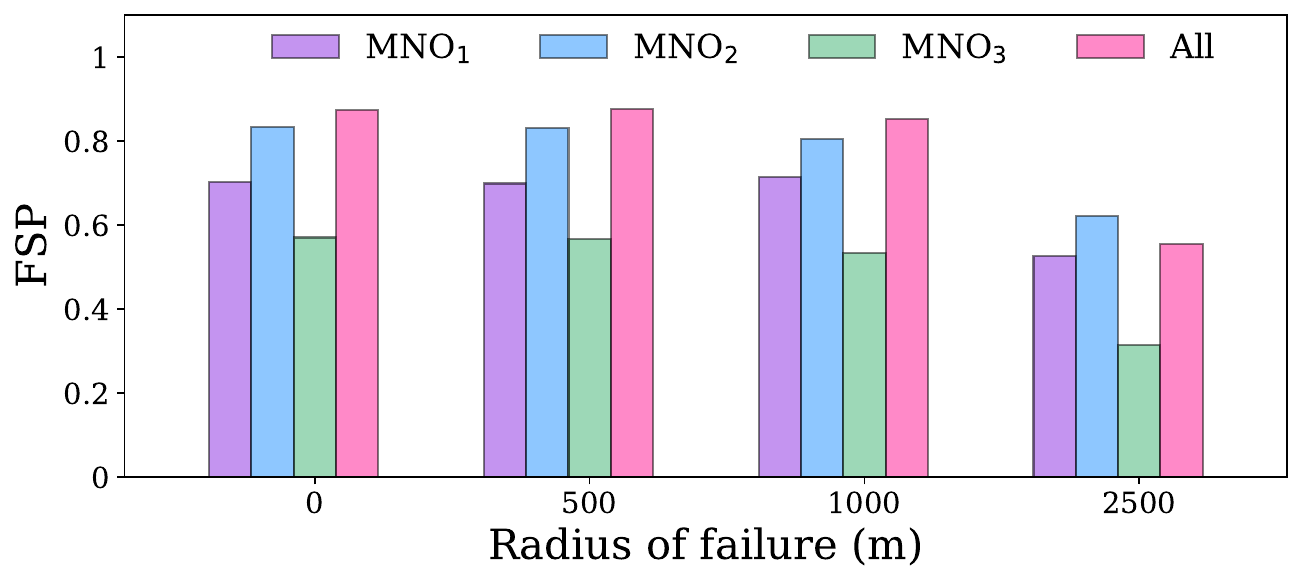}}\hfill
    \caption{FDP and FSP under correlated failures in Enschede, where all BSs within  $r_{\text{fail}}$ meters of the center fail. \label{fig:correlated_failure_nationalroaming}}
    
\end{figure*}

\textbf{Correlated failures:} Fig.~\ref{fig:correlated_failure_nationalroaming} plots the FDP and FSP with increasing radius of correlated failures. In every case, we simulate a failure in the center of the municipality and let all BSs within radius $r_\text{fail}$ m of the center fail. If $r_\text{fail} = 500$m, on average $1\%$ of the BSs in the region fail. For $r_\text{fail} = 100, 2500$ and $5000$m, this is respectively $9\%, 38\%$ and $65\%$ of the BSs.

Similar to the isolated failures in Fig.~\ref{fig:random_failure_nationalroaming}, Fig.~\ref{fig:correlated_failure_nationalroaming} suggests that the FDP and FSP are not affected for small regions of failure under correlated failures. However, for larger radii, we notice that national roaming does not result in the highest FSP and lowest FDP anymore~(e.g., when $r_{\text{fail}}=2500$ m), as \KPN performs better in terms of FDP and \TMobile in terms of FSP. 
We attribute this lower FSP to a higher number of users served by the remaining surviving BSs. Under national roaming, three times as many users are in the center of a region (the city center) compared to no national roaming. Moreover, due to the non-uniform deployment of BSs, the number of BSs that fail is relatively large compared to a single-MNO scenario. Hence, a disaster in this region has more impact when all users share the network compared to when every MNO uses its own network, as BSs on the border of the disaster region become more congested. Comparing Fig.~\ref{sfig:correlated_failure_NR_FSP} and Fig.~\ref{sfig:random_failure_NR_FSP}, correlated failures  causes a significantly lower satisfaction compared to isolated failures, e.g., for $r_\text{fail} = 2500$m, around $38\%$ of the BSs fail, but the FSP is lower than the FSP for $p_\text{iso} = 0.5$.

\begin{tcolorbox}[title=\textbf{Takeaway},colback=gray!20,colframe=white,arc=3mm,boxsep=3mm]
Our analysis shows that isolated failures do not lead to any significant FDP decrease as the MNOs have enough redundancy in terms of the BSs covering an area: even if 50$\%$ of the BSs in a region fails, most users still maintain the required minimum signal level for connectivity. On the contrary, the FSP  drastically decreases, as BSs become more congested and the effective user bandwidth decreases due to the increase in the number of users served by the surviving BSs. Correlated failures lead to a more significant impact compared to isolated failures, i.e., increase in FDP and decrease in FSP. However, in most cases, numerical analysis shows that national roaming has the potential to improve resilience. 
\end{tcolorbox}

\section{Conclusion}\label{sec:conclusion}
Due to the increasing importance of cellular networks in the operation of critical infrastructures, it is paramount to quantify the resilience of cellular networks and consequently to proactively develop strategies to mitigate potential risks. In this paper, we presented an approach to assess the resilience of a cellular network in case of various risks, e.g., isolated failures and correlated failures. 
Using the publicly-available data on cellular networks, population, and urbanity levels in the Netherlands, we showed the wide performance variance across different regions and operators in terms of the fraction of disconnected population and satisfied population. Moreover, we analyzed how much and where infrastructure sharing can offer benefits to each network operator {in terms of managing the specific regions or technologies which the MNOs share.}
Areas with lower resilience can be considered for deployment of new infrastructure or for inter-operator collaboration to benefit from the existing cellular network infrastructures. 
We believe that the presented model-based approach to investigate the resilience of a cellular network can be applied to other countries or areas. This would provide a way to compare infrastructure in different countries, and to learn best practices. 
Potential future work directions include understanding pricing mechanisms for operators to benefit from this operation mode and incorporating the core network into our models.

\end{document}